\documentclass[1p]{elsarticle}
\usepackage{adjustbox}
\usepackage{graphicx} 
\usepackage{tikz}
\usepackage{array}
\usepackage{booktabs}
\usepackage{framed}
\usepackage{listings}

\usepackage{subcaption}
\usepackage{pgfplots}
\usepackage{comment}
\usepackage{multirow}
\usepackage{rotating}
\usepackage{url}

\usepgfplotslibrary{groupplots}

\pgfplotsset{compat=newest}

\usetikzlibrary{arrows}
\usetikzlibrary{shapes}
\usetikzlibrary{positioning}
\usetikzlibrary{arrows.meta}
\usetikzlibrary{calc}
\usetikzlibrary{shapes.geometric}

\journal{}

\date{March 2025}

\begin{document}

\begin{frontmatter}

\title{Fully Automated Generation of Combinatorial Optimisation Systems Using Large Language Models}
\author[UoN]{Daniel Karapetyan}
\affiliation[UoN]{organization={University of Nottingham, School of Computer Science},
            country={United Kingdom}}

\begin{abstract}
    Over the last few decades, researchers have made considerable efforts to make decision support more accessible for small and medium enterprises by reducing the cost of designing, developing and maintaining automated decision support systems.
    However, due to the diversity of the underlying combinatorial optimisation problems, reusability of such systems has been limited; in most cases, expensive expertise has been required to implement bespoke software components.
    
    We explore the feasibility of fully automated generation of combinatorial optimisation systems using large language models (LLMs).
    An LLM will be responsible for interpreting the user-provided problem description in natural language and designing and implementing problem-specific software components.
    We discuss the principles of fully automated LLM-based optimisation system generation, and evaluate several proof-of-concept generators, comparing their performance on four optimisation problems.
\end{abstract}

\begin{keyword}
automated generation of algorithms \sep large language models \sep combinatorial optimisation \sep decision support systems
\end{keyword}

\end{frontmatter}

\section{Introduction}

Optimisation is one of the most ubiquitous mathematical tools, particularly crucial in decision support.
While some optimisation problems are mathematically trivial, others have far too many choices for a human to handle.
This gives rise to optimisation algorithms that can effectively search large spaces of solutions.

Over the last decades, the community of researchers and practitioners has rapidly advanced optimisation algorithms and optimisation systems (OSs), i.e.\ decision support/making systems based on them.
While optimisation is still computationally expensive, these OSs routinely achieve sufficiently good results in most applications.
The challenge, however, is that the development of OSs is often prohibitively expensive for individuals and small and medium enterprises, as it requires unique expertise and is time-consuming~\cite{Chouki2020}.
In other words, the bottleneck in many cases is not the algorithm performance but the cost of the OS development and maintenance.

These costs could potentially be reduced by reusing existing OSs, but, unfortunately, the reusability of OSs is limited due to the vast diversity of the optimisation problems~\cite{Du2022}.
This is particularly true for combinatorial optimisation problems, thus we focus on combinatorial optimisation in this paper.

Nevertheless, significant effort has been put into reducing the cost of decision support.
We can identify three main approaches:
\begin{enumerate}
    \item 
    Reuse of an OS designed to tackle a narrow family of optimisation problems.
    Such systems have user interfaces for customisation of the optimisation problem and feeding in the instance data.
    In complex cases, such interfaces require the user to write program code.
    This approach is practical only for relatively typical problems, as it needs many users to be financially viable.
    In this study, we focus on a generic approach that works on a wide range of optimisation problems.

    \item 
    Partial automation of the algorithm design process, often referred to as \emph{hyper-heuristics}~\cite{Burke2013}.
    In this approach, a human expert develops a set of problem-specific software components, and a problem-independent artificial intelligence (AI) framework forms an algorithm based on these components.

    \item 
    Reduction to a standard problem.
    Instead of designing a new algorithm for the problem in hand, one can model the problem using a formal language and then apply an off-the-shelf solver for that language.
    A widely used example of such a formal language is mixed-integer programming (MIP)\@. 
    The problem is modelled using MIP, and then the search for an efficient solution is delegated to an off-the-shelf MIP solver.
    This means, however, that an optimisation system based on an MIP solver needs additional routines to translate the problem into the MIP language and then translate the MIP solution back into the problem language.
    These routines are problem-specific and often non-trivial.
\end{enumerate}

Note that the second and third approaches still require a human expert to design and develop problem-specific components, including data structures and input/output routines.
Thus, the process involves \emph{the user} describing the decision problem and a human expert building the OS\@.
The aim of this paper is to explore the feasibility of full automation of OS generation to completely eliminate the need for a human expert.
The user will provide a description of the problem in a natural language.
Additionally, the user may be expected to provide a set of benchmark instances (training data).
Based on these inputs, a fully automated OS generator (FAOG) will produce an OS.

It should be noted that the design of an OS also requires mathematical modelling of the problem, i.e.\ describing it with mathematical primitives.
In this paper, we expect that the user is capable of problem modelling; the problem description provided by the user must be mathematically precise.
Our rationale is that modelling is often straightforward for simple optimisation problems; the number of people with the appropriate skills is higher than the number of people capable of both modelling a problem and developing an optimisation system.
However, we recognise that automation of modelling is an interesting direction for future research.

The main novelty of this study is the use of large language models (LLMs) to achieve full automation of OS design and development.
Modern LLMs are capable of producing computer code based on instructions in a natural language.
By carefully engineering LLM prompts with embeddings of the user-provided problem description and orchestrating the code composition and testing processes, we can expect to fully automate the generation of all the software components and, in turn, the entire OS.

Since the rise of LLMs, there have been several studies using LLMs for mathematical optimisation (most of them are not peer-reviewed yet).
Researchers explored the feasibility of using LLMs as the optimisation system~\cite{Guo2024,yang2024} or optimisation algorithm component~\cite{chen2024,Wang2024}.
In these studies, the instance data and, sometimes, the solution data are fed to the LLM and the LLM uses its intelligence to produce new problem solutions.
The key strength of this approach is its flexibility as LLMs can handle a wide range of problems and data representations.
However, this approach is limited due to the high computation cost of LLMs, restrictions on the prompt size and hallucinations.
There were also several attempts to use LLMs to design new metaheuristics~\cite{Michal2023,Zhong2024}.
These studies show the viability of using LLMs to generate algorithmic components based on textual descriptions however they are focused on continuous optimisation.
A recent study has also explored the feasbility of automated generation of linear and mixed-integer linear programs for combinatorial optimisation problems~\cite{Ahmaditeshnizi2023}.
To the best of our knowledge, there has been no research using LLMs to auto-generate heuristic combinatorial optimisation optimisation systems, which is the focus of this paper.

The main contributions of the paper are as follows:
\begin{itemize}
    \item 
    The novel concept of a FAOG employing an LLM to develop problem-specific software components.

    \item 
    A discussion of FAOG principles and strategies for tackling associated issues.

    \item 
    A combination of the LLM-based intelligence with the automated algorithm configuration.

    \item
    A library of problem descriptions for future research of automated OS generation.

    \item 
    A computational study with several proof-of-concept FAOGs.
\end{itemize}

The paper is structured as follows.
In Section~\ref{sec:approaches}, we discuss possible approaches to FAOG design.
Section~\ref{sec:design} gives details of our proof-of-concept FAOGs designed for computational experiments.
Section~\ref{sec:library} introduces a library of problem descriptions, and Section~\ref{sec:experiments} reports on the results of our computational study.
A discussion of our findings and concluding remarks are given in Section~\ref{sec:conclusions}.

\section{Principles of fully automated OS generation}
\label{sec:approaches}

We begin with setting our objectives in designing a FAOG and then discuss the overall architecture of FAOGs\@.
Finally, we focus on specific modules of a FAOG.

\subsection{FAOG design objectives}

The aim of a FAOG is to generate an OS based on a natural language description of the problem provided by the user.
Interpretation of the problem description requires natural language processing (NLP), making the OS generation inherently unreliable~\cite{Brcic2023}.
In other words, it is impossible to guarantee that the generator will always produce a working and correct OS.
Thus, our key objective is to maximise the success rate of the FAOG, i.e.\ the probability that, given an adequate problem description, it will produce a working and correct OS.

Observe that the design and development of high-performing optimisation algorithms traditionally require unique expertise and significant time investment.
While there have been many successful attempts to use artificial intelligence to develop optimisation algorithms, to the best of our knowledge, none of them has been shown to consistently deliver state-of-the-art performance across a wide range of optimisation algorithms without considerable interventions from human experts.
Thus, one may expect FAOGs to produce relatively inefficient algorithms; such algorithms may not be able to find sufficiently good solutions or may take too long.

Finally, a FAOG is only useful if it can produce an OS within a reasonable time.
As a reference point, one can use the time required for a human expert to design and develop an OS.

Considering the above observations, the key considerations when designing a FAOG are as follows (in the order of priority):
\begin{enumerate}
    \item 
    Success rate of the generation process: a FAOG should be able to generate a working OS for as many problems as possible.

    \item
    Quality of solutions found by the generated OS.

    \item 
    Time budget required for the generated OS to find reasonable solutions.

    \item 
    Time budget required for the OS generation.
\end{enumerate}

\tikzset{
    data/.style={draw, rectangle, rounded corners, minimum width=3cm, minimum height=0.5cm, align=center, fill=green!20!white},
    component/.style={draw, rectangle, minimum width=3cm, minimum height=0.5cm, align=center, fill=blue!20!white},
    generated component/.style={component, fill=green!20!white},
    arrow/.style={
        -{Stealth[scale=1.5]} 
    },
    bidirectional arrow/.style={{Stealth[scale=1.5]}-{Stealth[scale=1.5]}}
}

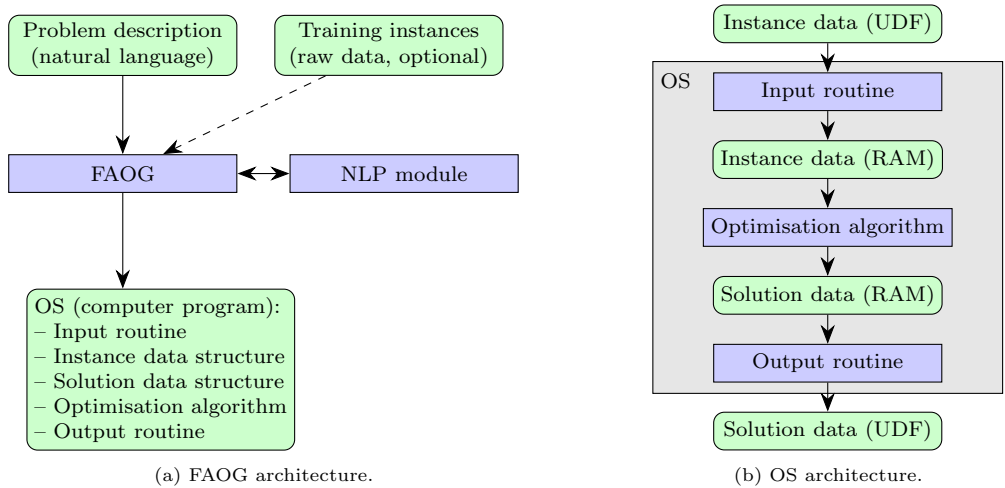
\begin{figure}[htb]
\centering

    \begin{subfigure}{0.55\textwidth}    
    \centering
    \begin{tikzpicture}
      \footnotesize
      \node[data] (problem description) at (-0.5, 0) {Problem description\\ (natural language)};
      \node[data] (instances) at (3, 0) {Training instances\\ (raw data, optional)};
      \node[component] (generator) at (-0.5, -1.7) {FAOG};
      \node[component] (nlp) at (3.2, -1.7) {NLP module};
      \node[data, align=left] (OS) at (0, -4.3) {OS (computer program):\\  
        -- Input routine \\
        -- Instance data structure \\
        -- Solution data structure \\
        -- Optimisation algorithm \\
        -- Output routine};
    
      \draw[arrow] (problem description) -- (generator);
      \draw[bidirectional arrow] (generator) -- (nlp);
      \draw[arrow] (generator.south) -- ($(generator)!(OS.north)!(generator)$);
      \draw[arrow, dashed] (instances) -- (generator);
    \end{tikzpicture}
    \caption{FAOG architecture.}
    \label{fig:os-generator-architecture}
    \end{subfigure}
    \hfill
    \begin{subfigure}{0.35\textwidth}
    \centering
    \begin{tikzpicture}
      \footnotesize
      \draw[fill=black!10!white, draw=black] (-2.3, -0.6) rectangle (2.3, -5);
      \node[draw=none] at (-2, -0.85) {OS};
      
      \node[data] (instance file) at (0, -0.1) {Instance data (UDF)};
      \node[component] (input routine) at (0, -1) {Input routine};
      \node[data] (instance data structure) at (0, -1.9) {Instance data (RAM)};
      \node[component] (optimisation algorithm) at (0, -2.8) {Optimisation algorithm};
      \node[data] (solution data structure) at (0, -3.7) {Solution data (RAM)};
      \node[component] (output routine) at (0, -4.6) {Output routine};
      \node[data] (solution file) at (0, -5.5) {Solution data (UDF)};
    
      \draw[arrow] (instance file) -- (input routine);
      \draw[arrow] (input routine) -- (instance data structure);
      \draw[arrow] (instance data structure) -- (optimisation algorithm);
      \draw[arrow] (optimisation algorithm) -- (solution data structure);
      \draw[arrow] (solution data structure) -- (output routine);
      \draw[arrow] (output routine) -- (solution file);
    \end{tikzpicture}
    \caption{OS architecture.}
    \label{fig:os-architecture}
    \end{subfigure}
    
    \caption{The architectures of a FAOG and the generated OS.}
    \label{fig:architecture}
\end{figure}

\subsection{FAOG architecture}

To implement an OS, one needs data structures to store the instance data and the solution data in random access memory (RAM), input/output routines to input the instance data in the user-defined format (UDF) and output the solution data in UDF, and an optimisation algorithm that takes the instance data and produces a solution, see Figure~\ref{fig:os-architecture}.
Figure~\ref{fig:os-generator-architecture} shows the architecture of a FAOG that takes the problem description in a natural language and uses an NLP module to interpret it.
Optionally, such a FAOG may also take a set of training instances.

During the OS generation process, the computer is expected to interpret the problem description in a natural language and produce computer code based on this description.
To the best of our knowledge, the only readily available tool capable of performing such tasks is LLM\@. 
An LLM-based FAOG can produce prompts that include the user-provided description of the problem and, for example, request the LLM to produce a Python function that calculates the objective value of a given solution.

While FAOGs can, hypothetically, generate algorithms of any type, we focus on heuristic algorithms in this study.
Heuristics are generally simpler than effective exact methods, and they can be composed of multiple simple algorithmic components.
We expect LLMs to work more reliably when the tasks are relatively simple.
Another reason for focusing on heuristic algorithms is that they are more likely to produce acceptable solutions within the available time budget.
For most business cases, a non-optimal solution produced within a time budget is preferable to an optimal solution produced well after it was needed.

To maximise the FAOG success rate, it is logical to use the divide and conquer approach: split the generation of various OS modules into several processes.
Each module can be tested independently, with mistakes being corrected before the rest of the system is generated.
This approach increases the probability of successful OS generation within a given number of attempts.

\subsection{Data structures and input/output routines generation}
\label{sec:data-structures}

An OS needs to store the instance data and the solution data, meaning that it needs appropriate data structures.
The specific data structures used by an OS depend on the problem and, thus, have to be generated each time a new OS being developed.

Note that some optimisation solvers use unified problem representations during the solution process (consider mixed-integer programming solvers).
However, this approach requires the original instance data to be translated into the unified representation, and then the solution to be translated into the domain-specific solution data structure.
Hence, the system still needs custom data structures for the instance and solution data.

The choice of the data structures and their implementation require intelligence; we rely on the LLM to supply this intelligence.
We can, however, assist the LLM by providing a clear and detailed set of instructions about the programming interfaces that must be implemented.
Among other functions, the instance data structure needs to support loading the instance data in UDF, and the solution data structure needs to support outputting the solution data in UDF\@.

\subsection{Optimisation algorithm generation}
\label{sec:algorithm-generator}

Once the data structures for storing instance data and solution data are composed, the FAOG needs to generate the optimisation algorithm, i.e. a routine that takes the instance data as the input and returns an efficient solution to the problem.
We propose the following classification of the optimisation algorithm generation approaches:

\begin{description}
    \item[Monolithic approach:] the LLM composes the entire optimisation algorithm.
    Specifically, it produces a function that takes the problem instance as the input and returns a solution to this problem.

    \item[Reduction-based approach:] the LLM composes routines to reduce (translate) the problem instance to another problem, run an external solver designed specifically for that problem and translate the solution back to the original problem.
    For example, the LLM-composed routine encodes the problem in MIP, feeds it to an off-the-shelf MIP solver, runs the MIP solver, and translates the MIP solution to the original problem solution.
    
    \item[Component-based approach:] the LLM composes simple algorithmic components that are used by a problem-independent framework as building blocks.
    Each component constructs or manipulates solutions.
    For example, a component may take a feasible solution and modify it while preserving its feasibility.
\end{description}

The above approaches can also be hybridised.
For example, some of the components in the component-based approach could use the reduction-based approach, or the problem-independent part of the component-based approach could also be produced by the LLM thus hybridising it with the monolithic approach.

\bigskip

An algorithm performing well on one family of problem instances might be relatively inefficient on another family of problem instances.
This gives rise to algorithm configuration and parameter tuning commonly used in optimisation algorithm design~\cite{Hamadi2012}.
The idea is to tweak the algorithm and/or its parameters so that it performs well on a set of training instances.
Then the expectation is that it will also perform well on the previously unseen but similar instances.

We distinguish FAOGs with and without offline training.
FAOGs with offline training require the user to provide a set of training instances in addition to the problem description and use these instances to improve the performance of the algorithm on these/similar instances.
FAOGs without offline training produce the algorithm solely based on the problem description.

There is also an adjacent concept of an adaptive algorithm; an adaptive algorithm performs learning during its run and adapts accordingly.
An algorithm generator of any type can potentially create an adaptive algorithm, however we do not explicitly explore generation of adaptive algorithms in this paper.

\section{FAOG design}
\label{sec:design}

To investigate the viability of fully automated OS generation, we built several proof-of-concept FAOGs based on the ideas discussed in Section~\ref{sec:approaches}.
Below, we provide the details of the design of our FAOGs.

\subsection{Problem description}

The input of a FAOG is the problem description produced by the user.
To help the user in structuring the problem description, the input is divided into several sections as detailed in Table~\ref{tab:problem-description-sections}.
This also makes it easy for the FAOG to include only the relevant information in each LLM prompt.

\begin{table}[htb]
    \centering
    \renewcommand{\arraystretch}{1.5} 
    \begin{tabular}{@{} >{\raggedright\arraybackslash}p{5em} p{12em} p{16em} @{}}
        \toprule
        Section & Section description & Example \\
        \midrule
        Input data 
            & Semantic description of the instance data.
            & Integer $n$.  Set $V$ of $n$ cities.  Positive integer cost $c(u, v)$ for each $u \in V$ and $v \in V$.\\
        Solution
            & Semantic description of the solution data.
            & A sequence of cities $(v_1, v_2, \ldots, v_n)$.\\
        Constraints
            & List of constraints that have to be satisfied for a solution to be feasible.
            & The solution should include exactly $n$ cities.  Cities cannot repeat.\\
        Instance file format
            & The description of the instance file format.
            & Text file.  The first line contains number $n$.  The next $n$ lines contain the $n \times n$ matrix giving the weights $c(u, v)$, where $u$ is the row index and $v$ is the column index.  Numbers are space-separated.\\
        Solution file format
            & The description of the solution file format.
            & Text file.  One line with n numbers giving the sequence of cities: $v_1$, $v_2$, \ldots, $v_n$.  The numbers are space-separated.  The first city has index 1.\\
        Example $i$ (for $i = 1, 2, \ldots$)
            & An example of an instance and a solution to it.
            & Instance: toy\_instance.txt\newline
Solution: toy\_solution1.txt\newline
Objective value: 8\\
        Training instances
            & A list of training instances.
            & instance1.txt\newline
            instance2.txt\\
        \bottomrule
    \end{tabular}
    \caption{The sections of the problem description file.}
    \label{tab:problem-description-sections}
\end{table}

\subsection{OS design}

Here, we discuss the design of the OS to be generated by our FAOG.

We chose Python as the programming language for this study.
While programs in Python are known to be significantly slower than programs in languages such as C++, it has several important advantages.
Most importantly, due to the vast amount of training data, LLMs are particularly good in Python.
Secondly, Python is a convenient language for dynamic code handling.
Thirdly, Python supports the object-oriented programming paradigm which is convenient for splitting the code into several relatively independent units.
Finally, Python is frequently used by the research community for distributing source codes as they are easy to understand and run on a wide range of hardware.

There are three main classes in our OS: instance class, solution class, and algorithm class (which in turn may use several other classes), see Figure~\ref{fig:uml}.
The instance class is responsible for reading the instance file and storing the instance data.
The solution class is responsible for creating a new random solution, storing solution data, calculating the objective value, testing the solution feasibility and saving the solution to a file.
The algorithm class is responsible for producing an efficient solution within a given time budget.

\tikzset{
    class/.style={
        rectangle,
        draw=black,
        fill=gray!10,
        rounded corners,
        align=left,
        minimum height=1cm,
        text width=5.5cm
    },
    inheritance/.style={draw, -{Triangle[open]}, thick},
    aggregation/.style={draw, -{Diamond[open]}, thick},
    composition/.style={draw, -{Diamond[fill=black]}, thick},
}

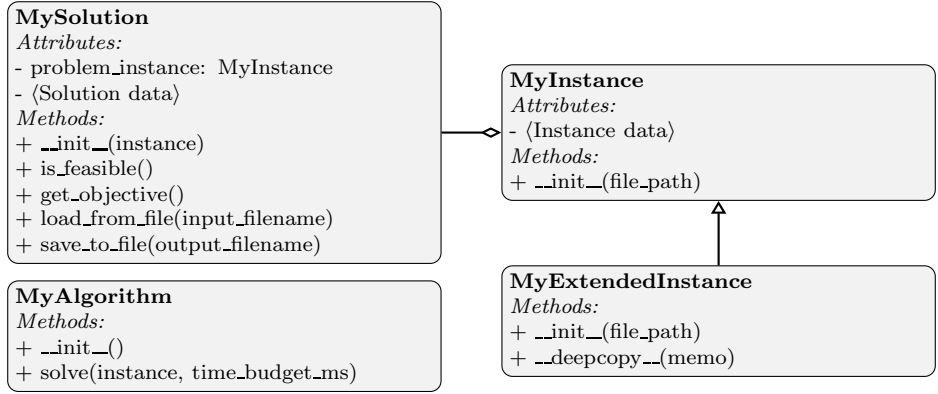
\begin{figure}[htb]
\centering
\begin{tikzpicture}[node distance=1.5cm]
\footnotesize

\node[class] (MySolution) {
    \textbf{MySolution}\\
    \textit{Attributes:}\\
    - problem\_instance: MyInstance\\
    - \textlangle Solution data\textrangle\\
    \textit{Methods:}\\
    + \_\_init\_\_(instance)\\
    + is\_feasible()\\
    + get\_objective()\\
    + load\_from\_file(input\_filename)\\
    + save\_to\_file(output\_filename)\\
};

\node[class, below of=MySolution, yshift=-1.2cm] (MyAlgorithm) {
    \textbf{MyAlgorithm}\\
    \textit{Methods:}\\
    + \_\_init\_\_()\\
    + solve(instance, time\_budget\_ms)
};

\node[class, right of=MySolution, xshift=5cm] (MyInstance) {
    \textbf{MyInstance}\\
    \textit{Attributes:}\\
    - \textlangle Instance data\textrangle\\
    \textit{Methods:}\\
    + \_\_init\_\_(file\_path)
};

\node[class, below of=MyInstance, yshift=-1cm] (MyExtendedInstance) {
    \textbf{MyExtendedInstance}\\
    \textit{Methods:}\\
    + \_\_init\_\_(file\_path)\\
    + \_\_deepcopy\_\_(memo)\\
};

\draw[inheritance] (MyExtendedInstance) -- (MyInstance);
\draw[aggregation] (MySolution.east) -- (MyInstance.west);

\end{tikzpicture}

\caption{UML diagram of the auto-generated OS\@.}
\label{fig:uml}
\end{figure}

\subsection{General principles of LLM prompt engineering}

LLMs are sensitive to prompts~\cite{leidinger2023}, meaning that good prompt engineering techniques can contribute to the success rate of a FAOG\@.
However, studies have shown that prompts optimised for one model may not perform well on another model, and more research is needed to develop our understanding of the effects of prompt engineering~\cite{leidinger2023, sclar2024}.
Considering that the FAOGs developed for this study are only proof-of-concept, we opt for simple prompts leaving more advanced prompt engineering and its analysis to future research.

Our FAOG divides the OS generation process into sub-tasks and executes multiple prompts followed by response validation and error correction.
This gives more opportunities to correct mistakes.

Each of our prompts requests the LLM to create a single class.
We aim to ensure that these classes are compatible with the rest of the OS (have the correct programming interfaces) and perform tasks as expected.
To achieve compatibility, our prompts include comprehensive lists of class methods and their signatures (lists of parameters with their types and the return type).
Additionally, some prompts ask to save data in instance variables with specific names.
The other implementation details are left to the `creativity' of the LLM\@.

In our experience, LLMs may not follow every instruction precisely.
Thus, to maximise the success rate of a FAOG, we reduce the reliance on the LLM where possible.
Once the problem-specific classes (MyInstance, MySolution and MyAlgorithm) are generated by the LLM, the FAOG adds problem-independent code to the OS implementation.
The challenge is to add code that will work with any LLM-generated code.
We found that the most reliable way of doing it is to extend LLM-generated classes overriding methods as appropriate.
Where this is impossible, the FAOG adds snippets of problem-independent code directly to the LLM-generated code.

Another lesson we learnt was to use identifiers that are clearly distinct from any common words.
For example, we use the class name `MyInstance' instead of `Instance' to avoid ambiguity.

\subsection{Instance and solution class generation}

Our OS generation process starts by creating a class for storing the instance data.
The instance class is also responsible for reading the instance data from a given file.

\begin{figure}[htb]

\begin{lstlisting}
### Problem description ###

Consider a combinatorial optimisation problem with the following input data.  <Input data>
    
A solution to the problem consists of the following.  <Solution>

The constraints are as follows.  <Constraints>

The objective function is as follows.  <Objective function>

### Instructions ###
Compose a Python class MyInstance with exactly one method: '__init__(self, file_path)'.  The __init__ method should open the file located at file_path, read the instance data from the file and save it into instance variables.  The file format is as follows. 
 <Instance file format>
    
Reply only with the code of MyInstance.  Include all the necessary import statements.  Do not include examples.
\end{lstlisting}
    
    \caption{Instance class prompt.}
    \label{fig:instance-prompt}
\end{figure}

The prompt for the code of the instance class is given in Figure~\ref{fig:instance-prompt}.
Once the LLM composes the instance class, we test it and respond to the LLM with the problem description if the test fails (see Section~\ref{sec:validation} for details).
We allow at most two attempts to correct any mistakes.

The solution class is responsible for creating a new random solution, storing solution data, calculating the objective value, and saving the solution to a file.
For testing purposes, it should also be able to verify that the solution is feasible and load a solution from a file.

\begin{figure}[htb]
\begin{lstlisting}
Produce a Python class MySolution with the following methods:
1. __init__(self, inst: MyInstance) that does the following:
- Saves the parameter 'inst' into an instance variable 'problem_instance'.  
- Composes a random solution to the problem specified by 'inst'.  The solution has to satisfy all the problem constraints.
- Saves the composed solution into instance variables.

2. is_feasible(self) -> bool that returns True if the solution satisfies all the problem constraints and False otherwise.  If the solution breaks some constraint, the method should also print an error message describing how exactly a constraint was broken.

3. get_objective(self) that calculates the objective value of the solution and returns it.  Assume that the solution satisfies all the constraints.

4. save_to_file(self, output_filename: str) that creates a file 'output_filename' and saves the solution to it.  The output file format is as follows.  <Solution file format>

5. load_from_file(self, input_filename: str) that opens the file 'input_filename' and loads the solution from it.  It should save the loaded solution in the current object.  The file format is the same as described in point 4.

Reply only with the code of MySolution.  Include all the necessary import statements.  Do not include examples.
\end{lstlisting}
    
    \caption{Solution class prompt.}
    \label{fig:solution-prompt}
\end{figure}

Our prompt for solution class is given in Figure~\ref{fig:solution-prompt}.
If the solution class fails its testing, we allow at most two attempts to correct mistakes.

\subsection{Algorithm generation}

We designed one algorithm generator for each of the three approaches discussed in Section~\ref{sec:algorithm-generator}.

\begin{figure}[htb]
\begin{lstlisting}
Compose a Python class MyAlgorithm with the following methods:
1. __init__(self).  The method should not do anything.

2. solve(self, instance: MyInstance, time_budget_ms: int) -> MySolution.  The method should find and return a heuristic solution to the problem instance specified in the parameter 'instance'.  The solution process should be terminated after 'time_budget_ms' milliseconds time.  Use <approach> approach.

Reply only with the code of MyAlgorithm.  Include all the necessary import statements.  Do not include examples.
\end{lstlisting}
    
    \caption{Monolithic algorithm class prompt.  The value of `approach' defines the type of the algorithm that the LLM is expected to generate.}
    \label{fig:algorithm-prompt}
\end{figure}

\paragraph{Monolithic algorithm generator.}
The LLM composes the entire algorithm class.
The prompt may instruct the LLM to use a specific algorithmic approach such as the simulated annealing or iterated local search, see Figure~\ref{fig:algorithm-prompt}.
This generator does not support offline training.

If the composed algorithm class fails the testing, we allow up to four attempts to correct the mistakes.

\begin{figure}[htb]
\begin{lstlisting}
Compose a Python class MyAlgorithm with the following methods:
1. __init__(self).  The method should not do anything.

2. solve(self, instance: MyInstance, time_budget_ms: int) -> MySolution.  The method should encode the problem as a mixed integer programming program and solve it using the Gurobi solver.  It should then create an instance of class MySolution and populate it with the solution found by Gurobi, even if Gurobi did not find an optimal solution.  The solution process should be terminated after 'time_budget_ms' milliseconds time.  If no solution is found within the time budget, return a random solution.

Reply only with the code of MyAlgorithm.  Include all the necessary import statements.  Do not include examples.
\end{lstlisting}
    
    \caption{MIP-based algorithm class prompt.}
    \label{fig:mip-algorithm-prompt}
\end{figure}
    
\paragraph{Reduction-based algorithm generator.}  The LLM composes the entire algorithm class, however it is instructed to use an external mixed integer programming solver to solve the problem, see Figure~\ref{fig:algorithm-prompt}.
Specifically, we chose to use the Gurobi mixed integer programming solver because it is widely used in the optimisation literature and has a convenient Python API\@.
This generator does not support offline training, however it produces adaptive algorithms considering that Gurobi includes adaptive search strategies.
    
If the composed algorithm class fails the testing, we allow up to four attempts to correct the mistakes.

\paragraph{Component-based algorithm generator.}  
The component-based approach requires a problem-independent framework to assemble an optimisation algorithm from several components.
We chose the Conditional Markov Chain Search (CMCS) as a framework designed specifically for the automation of metaheuristic generation~\cite{Karapetyan2017}.
The key strengths of CMCS are its simplicity, low overheads (making it a good choice if the algorithmic components take little time to run) and flexibility (for example, it can model several standard metaheuristics).

CMCS works as follows.
Let $\mathcal{H} = \{ H_1, H_2, \ldots, H_k \}$ be a set of algorithmic components.
Each component is seen by the framework as a black box that takes a solution as the input and modifies it according to the internal logic; the only requirement is that the component maintains the feasibility of the solution.
The search starts with a random solution.
At each iteration, CMCS selects the next component, applies it to the current solution and replaces the current solution with the updated one.
Additionally, CMCS keeps track of the best solution found so far.

The intelligence of CMCS lies in its logic for selecting the next component at each iteration.
The decision is based on two pieces of information: the last executed component and whether it improved the solution.
Thus, the entire component logic can be encoded with two transition matrices: one for the `success' of the last component and one for its `failure'.
These two matrices along with the component set form the CMCS configuration.

The components for the CMCS configuration are selected from a pool of algorithmic components.
To build the component pool, the FAOG requests the LLM to compose several mutation components, i.e.\ algorithmic components that apply random modifications to the given solution.
Mutation components are particularly easy to implement, hence our expectation is that an LLM will be capable of reliably composing a diverse set of mutation components.

Our prompting algorithm repeats the following steps until the number of mutation classes reaches 2, or the total number of attempts exceeds 10, whichever comes first:
\begin{enumerate}
    \item 
    Prompt the mutation class code (see Figure~\ref{fig:mutation-prompt}).
    Unless this is a prompt for the first mutation class, include the statement requesting the logic of the class to be different to the logic of the previous mutation classes.

    \item
    If the newly generated mutation class passes the testing, add it to the set of successful mutation classes.
    Otherwise, request a new implementation specifying the mistake.
    After two unsuccessful attempts to repair the current mutation class, move on to the next mutation class (for example, if class \verb!MyMutation1! could not pass the testing after two attempts to repair it, prompt a new mutation class \verb!MyMutation2!).
\end{enumerate}

\begin{figure}[htb]
\begin{lstlisting}
Compose Python class MyMutation<index> with the following methods:
1. __init__(self).  The method should not do anything.

2. apply(self, cur_solution: MySolution) -> None.  Assume that 'cur_solution' satisfies all the problem constraints.  The method should apply a random change to the 'cur_solution' object such that 'cur_solution' still satisfies all the problem constraints.  Do not use the is_feasible() method.

The logic of MyMutation<index> should be different to the logic of <list of previously generated mutation classes>.

Reply only with the code of MyMutation<index>.  Include all the necessary import statements.  Do not include examples.
\end{lstlisting}
    
    \caption{Mutation class prompt.  `Index' is the index of the mutation class that is being generated.}
    \label{fig:mutation-prompt}
\end{figure}

The solution and mutation classes give us functionality to compose random solutions, clone solutions and randomly modify them using the mutations.
We use this functionality to compose several additional components for the component pool:
\begin{description}
    \item[Strong mutation($n$): ]
    apply a mutation component $n$ times.
    For each mutation component, our FAOG adds one strong mutation: $n = 3$.

    \item[Hill-climber($n$): ]
    apply a mutation component and keep the new solution if it is better than the original one; otherwise backtrack to the original one.
    This process can be repeated $n$ times.
    For each mutation component, our FAOG adds three hill-climbers: $n = 10$, $n = 100$, and $n = 1000$.

    \item[Ruin \& Recreate: ]
    replace the current solution with a new random solution.
\end{description}

Having a pool of algorithmic components, we apply offline training to obtain an efficient CMCS configuration.
This involves selecting a subset of components and optimising the transition matrices.
We follow the training process proposed in~\cite{Karapetyan2018}.
Specifically, we restrict our CMCS to 2-component deterministic configurations, i.e.\ configurations where each row of each transition matrix includes exactly one non-zero element.
We enumerate all the `meaningful' 2-component deterministic configurations and test each of them on a set of training instances.
For each training instance, we rank the configurations based on the objective value, and then use the total rank of a configuration as a measure of its quality.
The training process returns the configuration with the best total rank.
For training, we randomly select five instances from the set of instances provided with the problem description.

\subsection{Validation and error correction}
\label{sec:validation}

LLMs are infamous for the so-called hallucination -- the phenomenon of returning of incorrect results~\cite{Farquhar2024}.
In our context, hallucinations can lead to issues of several types:
\begin{enumerate}
    \item 
    The LLM response has unexpected format (for example, it does not include any code).

    \item 
    The LLM returns code that contains compile-time errors.

    \item 
    The LLM-generated code does not support the expected programming interface; for example, a function signature is incorrect.

    \item 
    The LLM-generated code produces run-time errors.

    \item 
    The LLM-generated code produces infeasible solutions.

    \item 
    The LLM generates code that may never or practically never terminate (for example, the code produces random solutions until a feasible one is found but the probability of randomly creating a feasible solution is low, or the code includes a user prompt), or it does not respect the specified time budget.

    \item 
    The LLM-generated code produces feasible solutions within the given time budget but it is `unreasonable'.
    For example, the algorithm is restricted to exploring only a small subset of the search space, or it always returns a random solution.
\end{enumerate}

Issues 1--3 can often be avoided by careful prompt engineering and automated static testing: following each prompt, we automatically compile the code and check that it implements the expected classes and functions.

Issues 4--6 require dynamic testing.
Assuming the sample instances and solutions are provided by the user, we can run the following tests:
    reading instance files,
%
    reading solution files,
%
    correctly testing the feasibility of the provided solutions,
%
    correctly calculating the objective values of the provided solutions,
%
    running the algorithm and ensuring that it terminates within the specified time budget,
%
    testing feasibility of the newly produced solutions, and
%
    saving new solutions to files and reading them back.

Each time new code is produced by the LLM, the FAOG runs all available tests.
If any test fails, the FAOG attempts to fix the mistake instead of simply restarting the generation process; this increases the probability of a successful OS generation within a fixed number of attempts.
Specifically, it communicates the details of the mistakes to the LLM and prompts it to re-generate the last piece of code.
For example, if an exception is caught, it includes the information about the test that was running (e.g., `Failed to create an instance of MyInstance.'), the type of the exception, the text of the exception and the line where it occurred (the actual content of the line so that the LLM could relate to it).
If several attempts to fix a mistake fail, we exploit the stochasticity of LLMs and restart the generation process from scratch.
If three attempts to generate the OS fail, we assume that the FAOG failed.

\section{Library of problem descriptions}
\label{sec:library}

To evaluate the performance of our proof-of-concept FAOGs, we developed a library of problem descriptions, each provided with a set of benchmark instances and example solutions.

While the approach we chose for selecting optimisation problems and composing their descriptions is not systematic, we followed several principles to make the conclusions of this experimental research practically useful to a certain degree:
\begin{itemize}
    \item 
    Since a FAOG user is unlikely to have a computer science background, we avoid mentions of the standard problem names.

    \item 
    We aimed to include descriptions in various styles; some descriptions include mathematical notations while others are less formal.

    \item 
    The problem descriptions are unambiguous.

    \item
    The descriptions of the file formats are technically precise.
\end{itemize}

We recognise that our problem descriptions might not fully resemble such descriptions produced by users without expertise in optimisation and modelling.
In future, we would like to obtain problem descriptions from the target audience representatives to support future research on automated OS generation.

Our library of problem descriptions consists of four combinatorial optimisation problems:
\begin{enumerate}
    \item 
    Travelling Salesman Problem (TSP) -- widely used as a benchmark combinatorial optimisation problem.
    Our description of the problem includes mathematical notations and explicit formulas.

    We developed a simple file format for the instance data; in our format, the cost matrix is provided explicitly.

    For the benchmark instances, we selected all the TSPLIB~\cite{Reinelt1991} instances of sizes up to $n = 500$.
    For all those instances, the optimal solutions are known.

    \item 
    Generalised Travelling Salesman Problem (GTSP) -- a famous generalisation of TSP\@.
    Compared to TSP, GTSP is more complex in that it requires additional input data, and the solutions have additional constraints.

    We used the text-based file format from the GTSP Instance Library~\cite{Gutin2010}.

    For the benchmark instances, we selected all GTSP Instance Library instances of sizes up to $n = 500$.
    For all those instances, the optimal solutions are known~\cite{Fischetti1997}.

    \item 
    Assignment Problem (AP) -- a polynomially solvable problem with a wide range of applications.
    While it may seem like a bad choice for testing a system that generates heuristic algorithms, users may not be aware of AP being polynomially solvable, hence we argue that this choice is valid.
    Additionally, the AP is supposed to be solved efficiently by an MIP-based algorithm; our experiments will test if our FAOG is capable of exploiting this problem property.

    We developed a simple text-based file format for the instance data.

    We generated 10 pseudo-random instances of sizes $10, 20, \ldots, 100$. 
    Each weight is independently drawn at random from $\{ 0, 1, \ldots, 99 \}$ with a uniform distribution.

    \item 
    Exam Timetabling Problem (ETP) -- a relatively complex problem with constraints and objective function slightly different to what is typically used in the literature.
    Our aim is to test our system on a problem that LLMs have not seen before and that has a relatively complex structure.
    Specifically, we ask for an assignment of exams to time slots such that the minimum distance (in time slots) between exams for any student is maximised.
    Since this is a maximisation problem and our system only supports minimisation problems, our description of the objective function includes multiplying the minimum distance by $-1$.

    We generated 10 pseudo-random instances.
    To generate instance $i \in \{ 1, 2, \ldots, 10 \}$, we randomly choose the number of exams $n$ from $\{ 5, 6, \ldots, 5i + 5 \}$, the number of time slots from $\{ n, n + 1, \ldots, 2n \}$ and the number of students from $\{ 5, 6, \ldots, 5i + 5 \}$.
    Then, for each student, we choose the number of exams they take from $\{ 2, 3, 4 \}$ and then select those exams randomly.
    All the distributions are uniform.
\end{enumerate}

The detailed problem descriptions used as the FAOG inputs can be downloaded from \url{https://people.cs.nott.ac.uk/pszdk/problems_library.zip}.

\section{Experiments}
\label{sec:experiments}

To test our ideas, we conducted a series of experiments with our proof-of-concept FAOGs, see Section~\ref{sec:design}.
All our FAOGs share the parts that compose the data structures and input/output routines but differ in algorithm generation.
The list of FAOGs we experimented with is given in Table~\ref{tab:generators}.

\begin{table}
\centering
\renewcommand{\arraystretch}{1.5}
\footnotesize
\begin{tabular}{@{} lp{40em} @{}}
\toprule
Name & Description of the algorithm generator \\
\midrule
    Free & Monolithic algorithm generator; the choice of the algorithmic approach is left to the LLM. \\
    SA & Monolithic algorithm generator based on the simulated annealing metaheuristic, i.e.\ the algorithm generation prompt instructs the LLM to produce a simulated annealing algorithm. \\
    TS & Monolithic algorithm generator based on the tabu search metaheuristic. \\
    ILS & Monolithic algorithm generator based on the iterated local search metaheuristic. \\
    MIP & Reduction-based algorithm generator; the algorithm generation prompt instructs the LLM to use the Gurobi mixed integer programming solver to model and solve the problem. \\
    CMCS & Component-based algorithm generator based on the Conditional Markov Chain Search. \\
\bottomrule
\end{tabular}

\caption{List of FAOGs tested in this study.}
\label{tab:generators}
\end{table}

We used three LLMs in our experiments:
\begin{itemize}
    \item 
    \emph{ChatGPT 3.5} by OpenAI -- perhaps the most widely used LLM at the moment of writing the paper.

    \item 
    \emph{Gemini 1.5 Pro} by Google.

    \item 
    \emph{Llama 3.1 (405B)} by Meta -- one of the most widely used free LLMs.
\end{itemize}
Each FAOG was tested with all three LLMs, thus 18 OSs were produced in total.
We refer to the OS generated by a FAOG $x$ as `$x$ OS', for example, CMCS OS is the OS generated by the CMCS FAOG\@.

Table~\ref{tab:performance} shows the performance of the auto-generated OSs\@.
Each value is the gap to the best-known solution averaged over all the test instances:
\begin{equation}
\label{eq:gap}
\mathit{gap} = \frac{1}{n} \sum_{i=1}^n \frac{f_i - b_i}{b_i} \cdot 100\%,
\end{equation}
where $n$ is the number of test instances, $f_i$ is the objective value of the solution obtained by the OS for instance $i$, and $b_i$ is the objective value of the best-known solution for instance $i$.
The optimal solutions for TSP, AP and GTSP are known; for ETP, we use the best solution ever found by any of our algorithms during this study.
The time budget of each OS is 100~seconds per instance.

\newcommand{\best}[1]{\textbf{#1}}

\begin{table}
\centering
\footnotesize
    \begin{tabular}{llrrrrrr}
        \toprule
        LLM & Problem & Free & SA & TS & ILR & MIP & CMCS \\
        \midrule
        \multirow{4}{*}{\begin{turn}{90}ChatGPT\end{turn}}
& TSP & 1156.7\% & 102.0\% & 327.5\% & 451.6\% & 389.0\% & \best{16.9\%}\\
& AP & 131.3\% & 110.3\% & 68.6\% & 136.0\% & \best{0.0\%} & 85.1\%\\
& GTSP & 342.4\% & 327.1\% & 327.4\% & 340.5\% & -- & \best{29.1\%}\\
& ETP & 93.8\% & 100.0\% & 84.0\% & 100.0\% & 96.0\% & \best{30.8\%}\\
        \midrule
        \multirow{4}{*}{\begin{turn}{90}Gemeni\end{turn}}
& TSP & -- & (0\%) & -- & -- & -- & \best{25.2\%}\\
& AP & (0\%) & 1292.9\% & 1308.9\% & 152.2\% & \best{0.0\%} & 63.1\%\\
& GTSP & -- & \best{60.3\%} & -- & -- & (50\%) & --\\
& ETP & 100.0\% & \best{10.6\%} & 100.0\% & -- & 100.0\% & 32.8\%\\
        \midrule
        \multirow{4}{*}{\begin{turn}{90}Llama\end{turn}}
& TSP & 1189.7\% & 115.4\% & 735.9\% & (86\%) & (32\%) & \best{11.5\%}\\
& AP & 133.5\% & 110.6\% & 112.1\% & 62.0\% & \best{0.0\%} & 75.2\%\\
& GTSP & 60.7\% & 58.5\% & 54.0\% & 61.3\% & -- & \best{11.6\%}\\
& ETP & 100.0\% & 19.9\% & 81.6\% & 58.0\% & 100.0\% & \best{33.6\%}\\
        \bottomrule
    \end{tabular}

\caption{Performance of the auto-generated OSs.}
\label{tab:performance}
\end{table}

If an OS failed to solve some test instances within the given time budget, we report the percentage of the instances that were solved in brackets; for example, `(86\%)' means that the OS failed in 14\% of the tests.
If a FAOG failed to produce an OS, we use dash.

Based on the ChatGPT results in Table~\ref{tab:performance}, we conclude that most of the FAOGs reliably produce working OSs for every problem in the library.
The only exception is the MIP FAOG that failed to produce a working OS for GTSP\@.

The quality of solutions significantly depends on the FAOG used.
Letting the LLM choose the solution approach yields relatively bad results; the Free OSs showed poor performance across all the experiments.
Other monolithic generators (SA, TS and ILS) produced more effective OSs but it is hard to identify the winning approach based on our experiments.

The MIP OS predictably demonstrated excellent performance on AP; every LLM correctly formulated the problem, and Gurobi quickly solved every instance of AP to optimality.
However, the MIP FAOG was less successful on the other problems.
In many cases, the MIP FAOG failed to produce a working OS; in other cases, the Gurobi-based solver demonstrated poor results.
Indeed, straightforward MIP formulations of many problems including TSP are known to be inefficient.

The CMCS OSs demonstrated the most consistent performance across all the experiments, particularly on the NP-hard problems (TSP, GTSP and ETP)\@.

\bigskip

\tikzset{
    ChatGPT/.style={thick, red},
    Gemini/.style={thick, mark=square, blue},
    Llama/.style={thick, mark=x, green!80!black},
    cmcs/.style={red, mark=*},
    gurobi/.style={blue, mark=square},
    iteratedlocalsearch/.style={green!80!black, mark=triangle},
    tabusearch/.style={olive, mark=x},
    simulatedannealing/.style={gray, mark=|},
    None/.style={black, mark=o},
}

\newcommand{\timeplot}[4]{
    \addplot[thick, #2, #4] table[x=time, y=#1#2] {#3_time.txt};
}

\newcommand{\alltimeplots}[2]{
    \timeplot{#1}{None}{#2}{}
    \addlegendentry{Free}

    \timeplot{#1}{simulatedannealing}{#2}{}
    \addlegendentry{SA}

    \timeplot{#1}{tabusearch}{#2}{}
    \addlegendentry{TS}

    \timeplot{#1}{iteratedlocalsearch}{#2}{}
    \addlegendentry{ILS}
       
    \timeplot{#1}{gurobi}{#2}{}
    \addlegendentry{MIP}

    \timeplot{#1}{cmcs}{#2}{}
    \addlegendentry{CMCS}
}

\newcommand{\problemtimes}[3]
{
    \nextgroupplot[
        xmode = log,
        grid = both,
        title = {#2},
        width = 5.5cm,
        height = 5cm,
        legend to name=legend #1,
        #3
    ]
    
    \alltimeplots{ChatGPT}{#1}
}

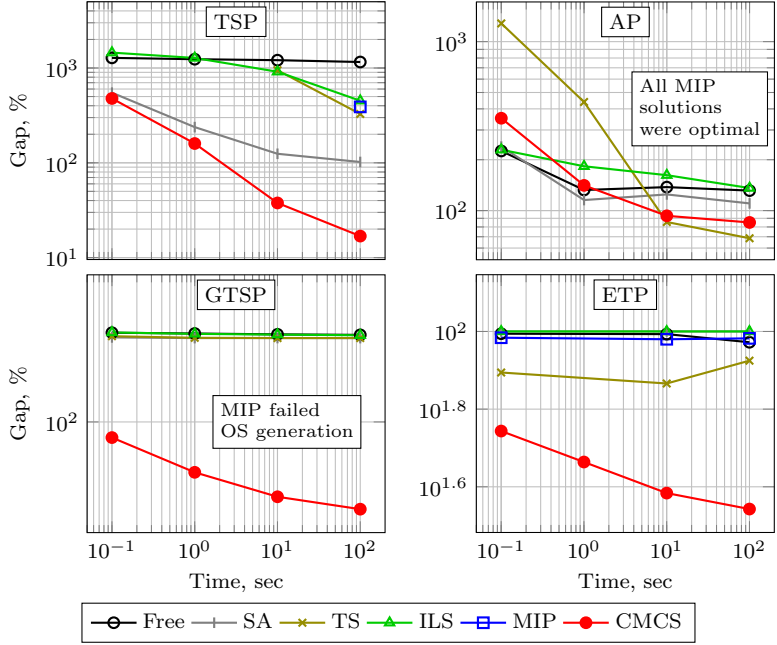
\begin{figure}[htb]
    \centering
    \footnotesize

    \begin{tikzpicture}
    \begin{groupplot}[
        group style={
            group size=2 by 2,
            horizontal sep=1.2cm,
            vertical sep=0.2cm,
        },    
        ymode=log,        
        legend cell align = left,
        legend style = {legend columns=6, /tikz/every column/.append style={column see]=1cm}},
        title style={
            at={(0.5,0.92)}, 
            anchor=north, 
            fill=white, 
            opacity=0.8, 
            text opacity=1, 
            draw=black
        }
    ]

    \problemtimes{tsp}{TSP}{xticklabels=\empty, ylabel={Gap, \%}, ymax=5000}
    \problemtimes{assignment}{AP}{xticklabels=\empty}
    \node[align=left, font=\scriptsize, draw=black, fill=white, fill opacity=0.8] at (25, 400) {All MIP\\ solutions\\ were optimal};
    \problemtimes{gtsp}{GTSP}{xlabel={Time, sec}, ylabel={Gap, \%}, ymax=800}
    \node[align=left, font=\scriptsize, draw=black, fill=white, fill opacity=0.8] at (13, 100) {MIP failed\\ OS generation};
    \problemtimes{examtimetabling}{ETP}{xlabel={Time, sec}, ymax=140}

    \end{groupplot}
    \end{tikzpicture}

    \pgfplotslegendfromname{legend tsp}

    \caption{Solution gap (\ref{eq:gap}) vs time budget per instance.  The FAOGs used in this experiment are based on ChatGPT.
    }
    \label{fig:chatgpt-perf}
\end{figure}

To understand how the solution time budget affects the quality of solutions, we conducted four experiments for each OS and problem instance, with time budgets 0.1~sec, 1~sec, 10~sec and 100~sec, respectively.
The LLM used in these experiments was ChatGPT\@.
The results are reported in Figure~\ref{fig:chatgpt-perf}.
Note that most of the auto-generated OSs are stochastic, meaning that the results of these experiments are noisy; an increased time budget can, in rare cases, give an increased solution gap.

The CMCS OSs clearly dominate all the other OSs across all the time budgets on the NP-hard problems.
The CMCS OS is also competitive on AP compared to the other metaheuristics\@.
The MIP OS solved all the AP instances to optimality even within the smallest time budget (the line is not visible because the y-axis is logarithmic).
However, MIP OS required a considerable time budget to produce any solutions for TSP, and the MIP FAOG failed the OS generation on GTSP.

Note that the only OSs in this paper that are capable of utilising multiple CPU cores are the MIP OSs; hypothetically, an LLM can produce parallelised code for any other FAOG but we have not observed such behaviours and do not believe that current LLMs are capable of reliably composing effective parallelised optimisation algorithms (partly because such algorithms are not common in the literature and hence the LLMs were not trained to produce them).
We also note that the multi-component approach makes it relatively easy to use concurrency even if the code produced by the LLM is sequential.

\newcommand{\comparellms}[3]
{
    \nextgroupplot[
        xmode = log,        
        grid = both,
        width = 5.5cm,
        height = 5cm,
        legend to name=legend #2,
        #3
    ]

    \timeplot{ChatGPT}{#1}{#2}{ChatGPT}
    \addlegendentry{ChatGPT}

    \timeplot{Gemini}{#1}{#2}{Gemini}
    \addlegendentry{Gemini}

    \timeplot{Llama}{#1}{#2}{Llama}
    \addlegendentry{LLama}
}

\begin{figure}[htb]
    \centering
    \footnotesize

    \begin{tikzpicture}
    \begin{groupplot}[
        group style={
            group size=2 by 2,
            horizontal sep=1.2cm,
            vertical sep=0.2cm,
        },    
        ymode=log,
        legend cell align = left,
        legend style = {legend columns=6, /tikz/every column/.append style={column see]=1cm}},
        title style={
            at={(0.5,0.92)}, 
            anchor=north, 
            fill=white, 
            opacity=0.8, 
            text opacity=1, 
            draw=black
        }
    ]

    \comparellms{cmcs}{tsp}{xticklabels=\empty, ylabel = {Gap, \%}, title={TSP}}
    \comparellms{cmcs}{assignment}{xticklabels=\empty, title={AP}}
    \comparellms{cmcs}{gtsp}{xlabel={Time, sec}, ylabel = {Gap, \%}, title={GTSP}}
    \comparellms{cmcs}{examtimetabling}{xlabel={Time, sec}, title={ETP}}

    \end{groupplot}
    \end{tikzpicture}

    \pgfplotslegendfromname{legend tsp}

    \caption{How the performance of the auto-generated CMCS depends on the LLM used.}
    \label{fig:cmcs-perf}
\end{figure}
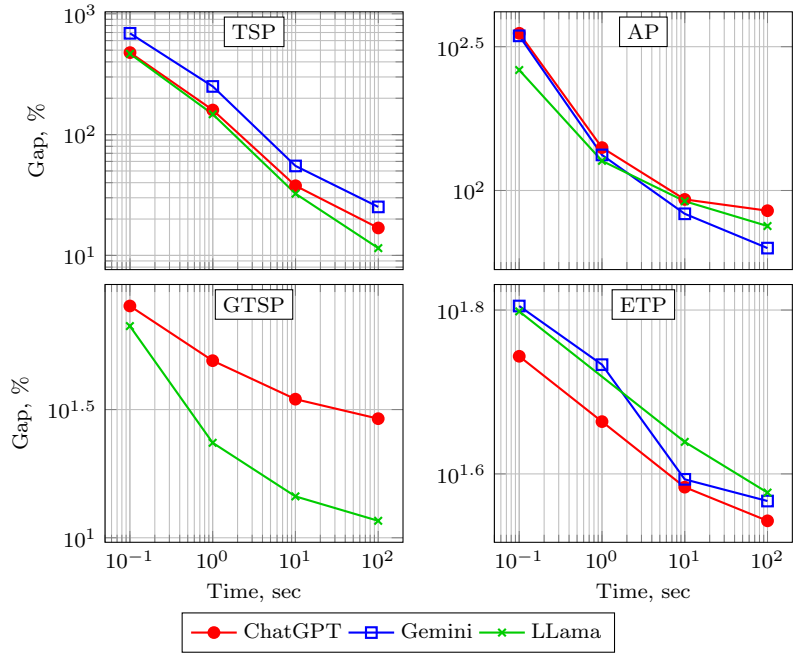

We also analysed how the performance of the generated OSs depends on the LLM used in the FAOG, see Figure~\ref{fig:cmcs-perf}.
We chose CMCS as the highest-performing FAOG for this experiment.
It is difficult to identify the winning LLM based on our limited results; with the exception of GTSP (arguably, the most sophisticated problem in our library), all the OSs performed similarly.
Our hypothesis is that CMCS is adaptive enough to perform well with a wide range of components; the offline learning mechanism compensates for the variability of the component quality.

An important observation is that the quality of solutions produced by the auto-generated OSs significantly improves with the time budget.
We conclude that the performance achieved by the CMCS FAOG might be sufficient for relatively small and simple problems.
We can also expect that further development of the generator and the use of more advanced LLMs will increase the success rate of the process and the performance of the OSs\@.

\bigskip

It is also important to discuss the running time of the FAOG\@.
For most of the approaches, OS generation consists of several LLM prompts and validation.
As a rule of thumb, the entire process takes in the order of one minute.
However, the CMCS generator also requires offline training that can take considerable time.
For our experiments, we restricted each run of CMCS OS during training to 1~second, and the set of instances used for training was limited to 5~instances.
In case of a two-component deterministic CMCS, there are around 300 `meaningful' configurations, thus the training takes around 25~CPU minutes.
Considering that CMCS training can be easily parallelised, this is a modest increase in the OS generation time.
However, training a less restricted CMCS could take much longer (hours or days) which may become a limiting factor in certain applications.

\section{Conclusions}
\label{sec:conclusions}

We explored the feasibility of full automation of combinatorial OS generation by utilising LLMs\@.
The input for such a generator is a description of a problem in a natural language and, optionally, a set of training instances.
The output is an OS, i.e.\ a software system that takes a problem instance data as input and outputs an efficient solution to this problem.
Such a generator will make decision support accessible to individuals and organisations that cannot currently afford it due to the cost of building OSs\@.

Within the study, we identified possible approaches to building FAOGs, developed proof-of-concept generators of several types and conducted a set of computational experiments to evaluate their performance.
For the computational experiments, we composed a library of problem descriptions and made it publicly available for future research.
We plan to maintain the library extending it with new problem descriptions, including descriptions produced by practitioners without expertise in optimisation.

Our experiments demonstrated that a FAOG utilising a modern LLM is capable of reliably producing OSs based on problem descriptions in English.
Among our proof-of-concept FAOGs, the most successful one uses the component-based approach with offline training.
In this approach, the LLM is responsible for interpreting the problem description and producing simple problem-specific algorithmic components whereas the problem-independent framework is responsible for building an efficient metaheuristic from those components.
Additionally, we found that the polynomially solvable problem from our library can be efficiently tackled with the reduction-based FAOG that utilises an MIP solver.
However, the reduction-based FAOG was less reliable and/or efficient on the other problems often producing incorrect reduction procedures and causing the OS to fail the automated testing.

Observe that the most successful FAOG is the only generator we tested that utilises offline training.
We hypothesise that offline training is crucial for generating efficient OS\@.
Note that problem descriptions usually do not include information about the instances that are expected to be solved in practice.
Such information is usually implicitly contained in the benchmark/training instances but may be unknown to the user.
Examples of such information are the number of isolated nodes in a graph, the size of the largest clique, the distribution of node degrees, etc.
An algorithm effective on one set of instances may not be effective on another set, hence training is important.
Traditionally, this training is conducted by human experts during the design and testing of algorithms.
To replicate this process, a FAOG needs to implement formal offline training.

Another argument for requesting the user to provide training instances along with the problem description is the validation of the code produced by the LLM\@.
Without such tests, there is a high chance of producing a disfunctional OS that, either, terminates without returning a solution or may return infeasible solutions.
For most of the applications, it is preferable to catch such issues before system deployment, i.e.\ during the generation process.

We also observe that the reduction-based and component-based FAOGs delegate more tasks to specialised AI compared to the monolithic generators; we can expect a specialised AI to perform better, thus giving an edge to these approaches.

In our experiments, the performance of the auto-generated OSs for the NP-hard problems was low compared to the state-of-the-art algorithms found in the literature; modestly sized instances required considerable time budgets to find solutions of acceptable quality.
However, we argue that even the OSs generated by our proof-of-concept FAOGs might be sufficient for some applications; considering the motivation for this study, it is more appropriate to compare the auto-generated OSs to human operators solving the problem.
We also believe that further development of the FAOGs will lead to significant performance improvements.

\subsection{Future work}

While our experiments demonstrated an acceptable success rate of the FAOGs, further improvements will enable generation of OSs for more complex problems.
Future research will identify more effective LLM prompting, code validation, and error correction.
For example, our implementation uses a single `conversation' with an LLM; every prompt includes the entire conversation history (unless a complete restart is performed).
A potentially more efficient approach will break up this conversation into multiple conversations, with the relevant code fragments embedded into the prompt.
Also, some classes such as MySolution can be further divided.
For example, the algorithm constructing a random solution can be separated from the MySolution class.
Finally, prompt engineering techniques such as chain-of-thought may improve the quality of the LLM responses~\cite{Espejel2024,Murr2023,sahoo2024}.

No matter how advanced a FAOG is, it needs a precise description of the problem to perform well.
Since some users will not have expertise in optimisation, they might need assistance in composing the problem description.
The OS generation process can be made interactive, with the FAOG providing feedback about the problem description (for example, highlighting ambiguities) and requesting clarifications.
It can also summarise, in a natural language, the algorithmic ideas it produced to let the user suggest new algorithmic ideas or identify inaccuracies in the problem description.

With regards to the performance of the auto-generated OSs, the following are a few areas where, we believe, it is possible to get significant gains:
\begin{itemize}
    \item
    A mutation component often takes $O(1)$ time, however the CMCS framework re-calculates the objective value each time a mutation (or any other component) is applied.
    For a problem of size $n$, calculation of the objective value usually takes $O(n)$ time; as a result, solution evaluations could consume most of the time budget.
    State-of-the-art algorithms often employ \emph{incremental evaluation}; a component does not only modify the solution but also evaluates how this modification affects the objective value; this often takes only $O(1)$ time.
    Incremental evaluation will require an adjustment to current programming interfaces, more complex prompts, and additional automated tests.
	
    \item
    At the moment, the CMCS FAOG prompts the LLM to compose random construction heuristic and mutation components whereas hill-climbers are later produced based on the mutations.
    While the current approach is sufficient to build an OS, the LLM might be able to compose more efficient hill-climbers.
	
    \item
    The auto-generated OSs can utilise all the available CPU cores to further improve the performance.
    This can be achieved by running one instance of the algorithm on each CPU core, or a more sophisticated strategy such as a memetic algorithm can be employed.
	
    \item	
    We found that the monolithic algorithm generator often produced reasonable algorithms which, however performed poorly due to the lack of parameter tuning.
    It is possible to enhance the monolithic FAOG by introducing offline parameter tuning.
\end{itemize}

Finally, a more advanced CMCS training procedure will allow finding more complex and effective CMCS configurations.

\bibliographystyle{elsarticle-num}

\begin{thebibliography}{10}
\expandafter\ifx\csname url\endcsname\relax
  \def\url#1{\texttt{#1}}\fi
\expandafter\ifx\csname urlprefix\endcsname\relax\def\urlprefix{URL }\fi
\expandafter\ifx\csname href\endcsname\relax
  \def\href#1#2{#2} \def\path#1{#1}\fi

\bibitem{Chouki2020}
M.~Chouki, M.~Talea, C.~Okar, R.~Chroqui, Barriers to information technology adoption within small and medium enterprises: A systematic literature review, International Journal of Innovation and Technology Management 17~(1) (2020).

\bibitem{Du2022}
D.-Z. Du, P.~Pardalos, X.~Hu, W.~Wu, Introduction to Combinatorial Optimization, Springer International Publishing, Cham, 2022.

\bibitem{Burke2013}
E.~Burke, M.~Gendreau, M.~Hyde, G.~Kendall, G.~Ocha, E.~\"Ozcan, R.~Qu, Hyper-heuristics: a survey of the state of the art, Journal of the Operational Research Society 64 (2013).

\bibitem{Guo2024}
P.-F. Guo, Y.-H. Chen, Y.-D. Tsai, S.-D. Lin, Towards optimizing with large language models (2024).
\newblock \href {http://arxiv.org/abs/2310.05204} {\path{arXiv:2310.05204}}.

\bibitem{yang2024}
C.~Yang, X.~Wang, Y.~Lu, H.~Liu, Q.~V. Le, D.~Zhou, X.~Chen, Large language models as optimizers (2024).
\newblock \href {http://arxiv.org/abs/2309.03409} {\path{arXiv:2309.03409}}.

\bibitem{chen2024}
Y.~Chen, Y.~Li, B.~Ding, J.~Zhou, On the design and analysis of {LLM}-based algorithms (2024).
\newblock \href {http://arxiv.org/abs/2407.14788} {\path{arXiv:2407.14788}}.

\bibitem{Wang2024}
Z.~Wang, S.~Liu, J.~Chen, K.~C. Tan, Large language model-aided evolutionary search for constrained multiobjective optimization, in: Advanced Intelligent Computing Technology and Applications: 20th International Conference, ICIC 2024, Tianjin, China, August 5–8, 2024, Proceedings, Part II, Springer-Verlag, Berlin, Heidelberg, 2024, p. 218–230.

\bibitem{Michal2023}
M.~Pluhacek, A.~Kazikova, T.~Kadavy, A.~Viktorin, R.~Senkerik, Leveraging large language models for the generation of novel metaheuristic optimization algorithms, in: Proceedings of the Companion Conference on Genetic and Evolutionary Computation, GECCO '23 Companion, Association for Computing Machinery, New York, NY, USA, 2023, p. 1812–1820.

\bibitem{Zhong2024}
R.~Zhong, Y.~Xu, C.~Zhang, J.~Yu, Leveraging large language model to generate a novel metaheuristic algorithm with crispe framework, Cluster Computing 27~(10) (2024) 13835–13869.

\bibitem{Ahmaditeshnizi2023}
A.~AhmadiTeshnizi, W.~Gao, M.~Udell, Optimus: Optimization modeling using mip solvers and large language models (2023).
\newblock \href {http://arxiv.org/abs/2310.06116} {\path{arXiv:2310.06116}}.

\bibitem{Brcic2023}
M.~Brcic, R.~V. Yampolskiy, Impossibility results in {AI}: A survey, ACM Comput. Surv. 56~(1) (Aug. 2023).

\bibitem{Hamadi2012}
Y.~Hamadi, {\'{E}}.~Monfroy, F.~Saubion (Eds.), Autonomous Search, Springer, 2012.

\bibitem{leidinger2023}
A.~Leidinger, R.~van Rooij, E.~Shutova, The language of prompting: What linguistic properties make a prompt successful? (2023).
\newblock \href {http://arxiv.org/abs/2311.01967} {\path{arXiv:2311.01967}}.

\bibitem{sclar2024}
M.~Sclar, Y.~Choi, Y.~Tsvetkov, A.~Suhr, Quantifying language models' sensitivity to spurious features in prompt design or: How i learned to start worrying about prompt formatting (2024).
\newblock \href {http://arxiv.org/abs/2310.11324} {\path{arXiv:2310.11324}}.

\bibitem{Karapetyan2017}
D.~Karapetyan, A.~P. Punnen, A.~J. Parkes, Markov chain methods for the bipartite boolean quadratic programming problem, European Journal of Operational Research 260~(2) (2017) 494--506.

\bibitem{Karapetyan2018}
D.~Karapetyan, B.~Goldengorin, Conditional markov chain search for the simple plant location problem improves upper bounds on twelve korkel-ghosh instances, in: B.~Goldengorin (Ed.), Optimization Problems in Graph Theory, Springer, 2018, pp. 123--147.

\bibitem{Farquhar2024}
S.~Farquhar, J.~Kossen, L.~Kuhn, Y.~Gal, Detecting hallucinations in large language models using semantic entropy, Nature 630~(8017) (2024) 625--630.

\bibitem{Reinelt1991}
G.~Reinelt, {TSPLIB} -- a traveling salesman problem library., INFORMS J. Comput. 3~(4) (1991) 376--384.

\bibitem{Gutin2010}
G.~Gutin, D.~Karapetyan, A memetic algorithm for the generalized traveling salesman problem, Natural Computing: An International Journal 9~(1) (2010) 47–60.

\bibitem{Fischetti1997}
M.~Fischetti, J.~J.~S. González, P.~Toth, A branch-and-cut algorithm for the symmetric generalized traveling salesman problem, Operations Research 45~(3) (1997) 378--394.

\bibitem{Espejel2024}
J.~L. Espejel, M.~S.~Y. Alassan, M.~Bouhandi, W.~Dahhane, E.~H. Ettifouri, Low-cost language models: Survey and performance evaluation on python code generation, arXiv preprint (2024).
\newblock \href {http://arxiv.org/abs/2404.11160} {\path{arXiv:2404.11160}}.

\bibitem{Murr2023}
L.~Murr, M.~Grainger, D.~Gao, Testing llms on code generation with varying levels of prompt specificity, arXiv preprint (2023).
\newblock \href {http://arxiv.org/abs/2311.07599} {\path{arXiv:2311.07599}}.

\bibitem{sahoo2024}
P.~Sahoo, A.~K. Singh, S.~Saha, V.~Jain, S.~Mondal, A.~Chadha, A systematic survey of prompt engineering in large language models: Techniques and applications (2024).
\newblock \href {http://arxiv.org/abs/2402.07927} {\path{arXiv:2402.07927}}.

\end{thebibliography}

\end{document}